\pdfoutput=1
\documentclass[sigconf]{acmart}

\usepackage{booktabs}
\usepackage{graphicx}
\usepackage{subcaption}
\usepackage{balance}
\usepackage{adjustbox}
\usepackage{footmisc}

\setcopyright{none}

\renewcommand\footnotetextcopyrightpermission[1]{}
\begin{document}
\title{Predict your Click-out: Modeling User-Item Interactions and Session Actions in an Ensemble Learning Fashion}

\author{Andrea Fiandro}
\affiliation{Politecnico di Torino}
\email{andrea.fiandro@studenti.polito.it}

\author{Giorgio Crepaldi}
\affiliation{Politecnico di Torino}
\email{giorgio.crepaldi@studenti.polito.it}

\author{Diego Monti}
\orcid{0000-0002-3821-5379}
\affiliation{Politecnico di Torino}
\email{diego.monti@polito.it}

\author{Giuseppe Rizzo}
\orcid{0000-0003-0083-813X}
\affiliation{LINKS Foundation}
\email{giuseppe.rizzo@linksfoundation.com}

\author{Maurizio Morisio}
\orcid{0000-0001-7362-906X}
\affiliation{Politecnico di Torino}
\email{maurizio.morisio@polito.it}

\begin{abstract}
This paper describes the solution of the POLINKS team to the RecSys Challenge 2019 that focuses on the task of predicting the last click-out in a session-based interaction. We propose an ensemble approach comprising a matrix factorization for modeling the interaction user-item, and a session-aware learning model implemented with a recurrent neural network.
This method appears to be effective in predicting the last click-out scoring a 0.60277 of Mean Reciprocal Rank on the local test set.
\end{abstract}

\maketitle

\section{Introduction}
\label{sec:introduction}

In recent years, recommender systems improved both user experience and company profits in many fields, ranging from e-commerce to music and video streaming services~\cite{Xiao2007}. Similarly, the tourism industry has focused on improving customer experience to encourage travellers to use a booking platform again~\cite{ricci2002travel}. 

In this paper, we describe the solution of the POLINKS team to the RecSys Challenge 2019.\footnote{\url{http://www.recsyschallenge.com/2019/}} The purpose of the challenge is to predict the hotel target of the last \textit{clickout action} of a session because it is the one that leads the user to the hotel company site and it is monetized.
The implementation of our method is publicly available at \url{https://github.com/D2KLab/touringrec}. Our approach consists of an ensemble of two different solutions: Matrix Factorization (MF) and Recurrent Neural Network (RNN)~\cite{ensemble2018}. In our vision, the value of the ensemble could reside in the complementarity of the methods. The MF directly associates a user to the hotels she interacted with, while the RNN extracts latent features from the sequence of the interactions. Our approach achieved a final MRR of $0.60277$ on the local test set.

The remainder of this paper is structured as follows: Section~\ref{sec:approach} describes our approach while, in Section~\ref{sec:results}, we present the results and we analyze them. Finally, we discuss the outcomes and limitations of our work in Section~\ref{sec:conclusion}.

\section{Approach}
\label{sec:approach}

Our approach is a multi-stage process composed of a dataset manipulation phase, followed by a feature extraction step, which is done separately on two subsets created from the original one.

Every action has an \textit{action type} field and a \textit{reference} field containing the action target. We decide to filter the actions in order to keep only those related to hotels\footnote{Hotel related actions: \{`Interaction item rating', `Interaction item info', `Clickout item', `Interaction item image', `Interaction item deal', `Search for item'\}.} as they are the most relevant ones for the prediction. We also keep track of the \textit{impression\_list} field of the clickout actions. The \textit{impression\_list} contains all hotels shown to the user. Our approach requires to initially split the dataset provided by Trivago in two subsets.

The first subset is related to cold-start sessions, i.e. sessions of length $len = 1$. The lack of previous actions makes them impossible to be processed by a machine learning method. This leads us to exploit an approach based on the \textit{impression\_list} field of the action to be predicted, as it is the only relevant information we have. More precisely, we simply use the \textit{impression\_list} as the recommendation list, without any change to it.

The second subset contains sessions of length $len > 1$. In the following sections we describe the two methods that we considered.

In order to evaluate the performance of our solutions locally, we randomly split the Trivago official training set in two sets: \textit{local training set} (80\%), and \textit{local test set} (20\%).
Those two dataset are created following the same structure of the original split provided by the challenge, avoiding the separation of actions belonging to the same session. To obtain a local test set similar to the original one we nullify the last clickout reference of each session.

\begin{figure}
    \centering
    \includegraphics[width=\linewidth]{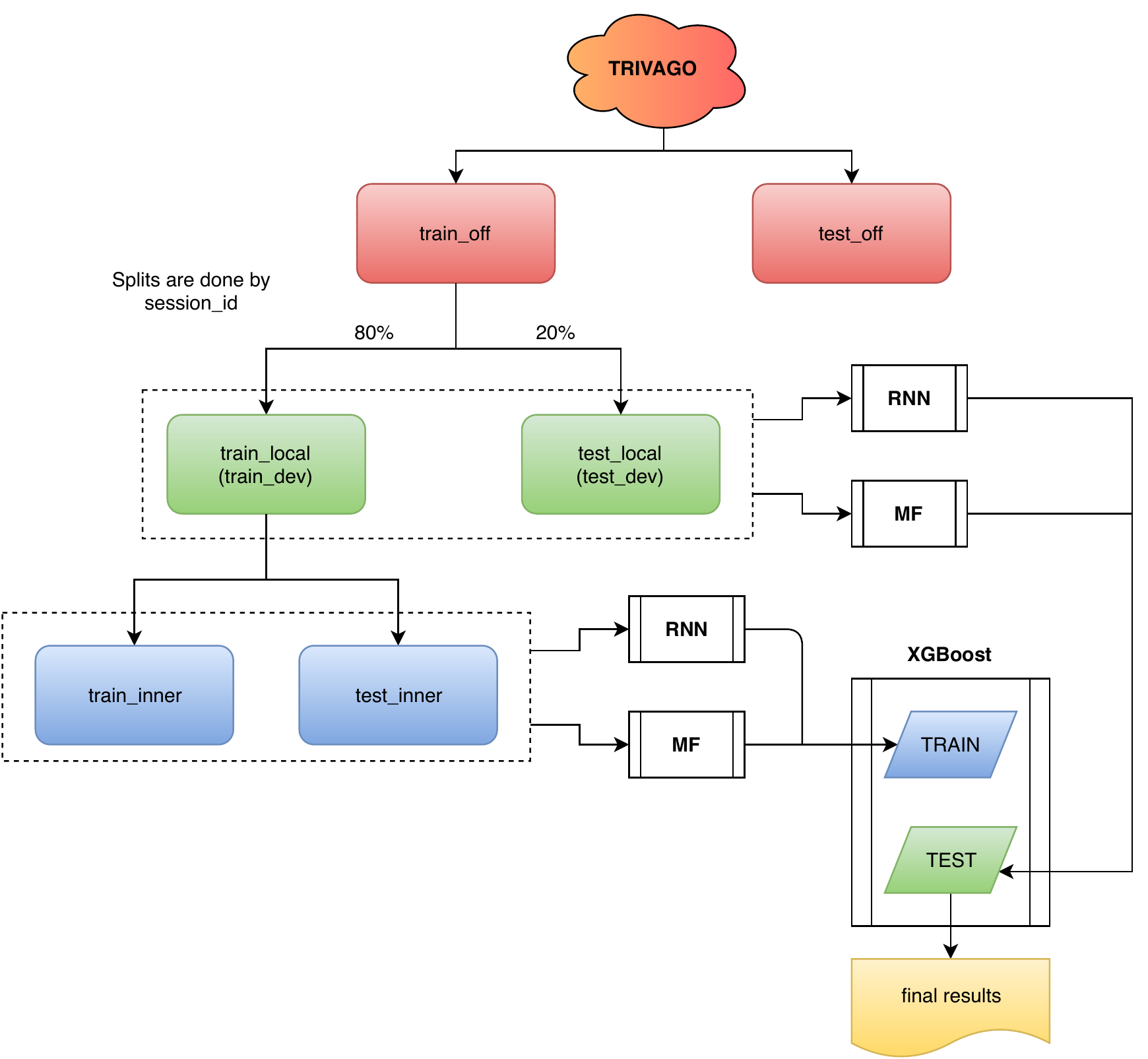}
    \caption{Scheme of dataset split and usage for the overall solution.}
    \label{fig:split_struct}
\end{figure}

We repeat the split on the \textit{local training set} to obtain a \textit{inner training set} and a \textit{inner test set}. The latter will be used to train and test the XGBoost model, which will give the final results based on the \textit{local} dataset. We decide to ignore the Trivago official test set in this experiment because the lack of labels for the fields to be predicted makes impossible to evaluate the results. The full split structure is shown in Figure~\ref{fig:split_struct}.

\subsection{Ensemble}
\label{sec:ensemble}

Our ensemble approach is used to integrate the results of the recurrent neural network with the matrix factorization model.
The ensemble works by giving more importance (more weight) to the top ranked hotel of each solution, like in a Borda count election.\footnote{\url{https://en.wikipedia.org/wiki/Borda_count}}

\subsection{Matrix Factorization}
\label{sec:mf}

Matrix Factorization (MF) is one of the most used algorithm in the context of recommender systems~\cite{koren2009matrix}. 
However, the main issue of this approach is the problem definition: it is important to choose a computationally feasible encoding. In the following sections, we will describe the generation strategy of the input matrices that we defined, the features extracted by the model and the gradient boosting algorithm.

\begin{figure}
    \includegraphics[width=\linewidth]{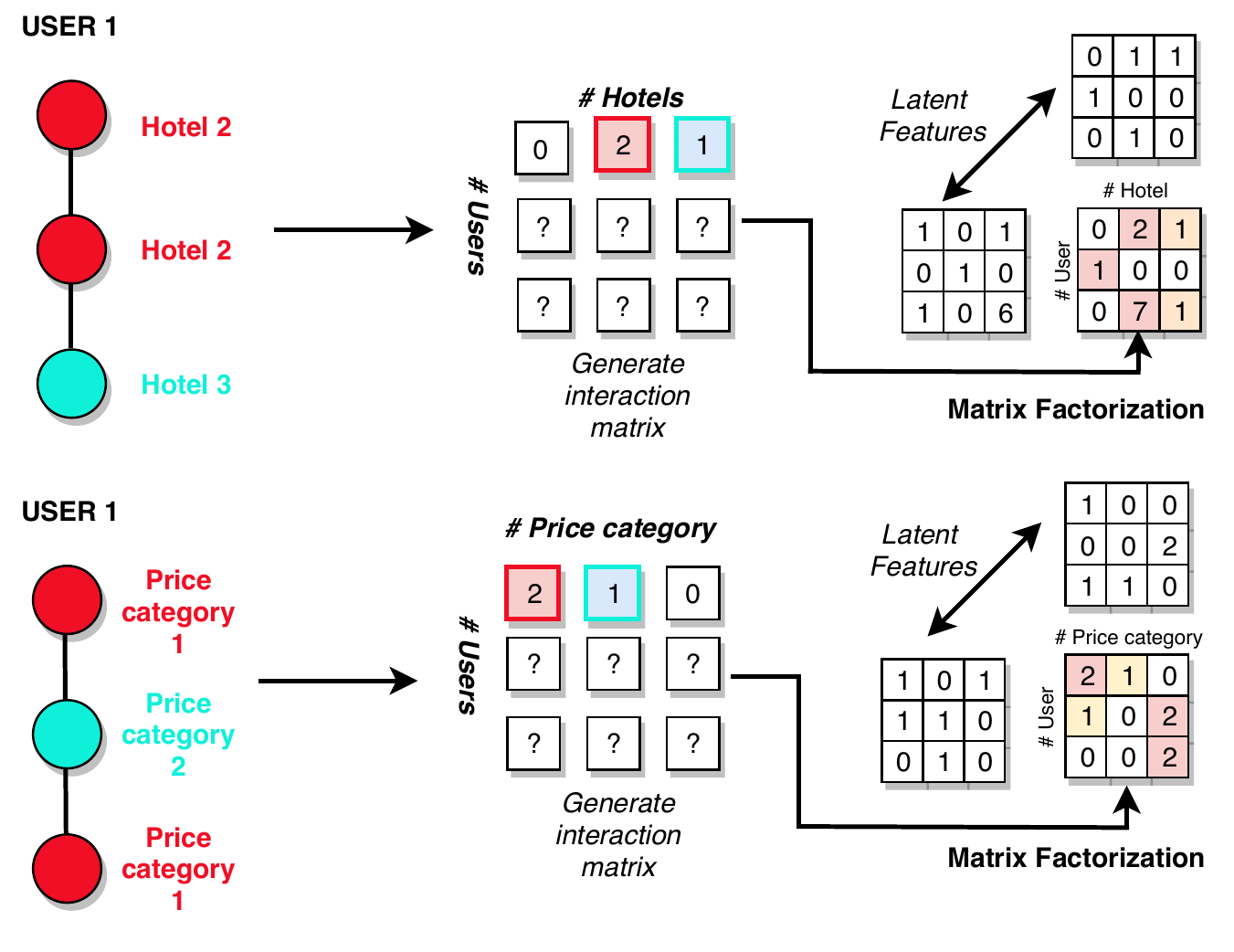}
    \caption{Interaction matrix generation.}
    \label{fig:mf}
\end{figure}

\subsubsection{Input matrices}
\label{sec:mf_input}
In our MF model, we consider two interaction matrices: the user-hotel and the user-price category ones.

The first matrix is extensively used in literature. However, we need to choose which kind of action can be considered as an interaction. We decide to use the actions that can be associated, distinctively, with a user-hotel pair, as introduced in Section~\ref{sec:approach}.

The second matrix uses the same interaction actions, but it takes into account only the price category of the hotel selected by the user. This kind of approach was used only in the cold start scenario, to predict a score for the user-hotel pair where the hotel was not previously seen by the user.

\subsubsection{Feature extraction}
\label{sec:mf_feature}

Our model learns the latent representations in a high dimensional space for user and hotels. When multiplied together, these representations gave us a score for every hotel for a given user. For example, we take a row of the user latent feature matrix $U_1 = (u_{f1}, u_{f2}, \dotsc, u_{fn})$ and a column of the hotel latent feature matrix.

\begin{equation*}
H_1 = \begin{pmatrix}
h_{f1}\\ 
h_{f2}\\ 
...\\
h_{fn}
\end{pmatrix}
\end{equation*}

In order to get the score $S_{11}$ for the pair ($U_1$, $H_1$), it is necessary to compute the dot product $U_{1} \cdot H_{1} = S_{11}$.

The scores are obtained using a trained LightFM model~\cite{Kula2015}, which implements the MF algorithm. We do not take into account only the score given by the MF to generate the prediction, but we use it as a feature inside a XGBoost model (along with \textit{user bias} and \textit{item bias}), as described in the following section.

\begin{figure}
    \includegraphics[width=\linewidth]{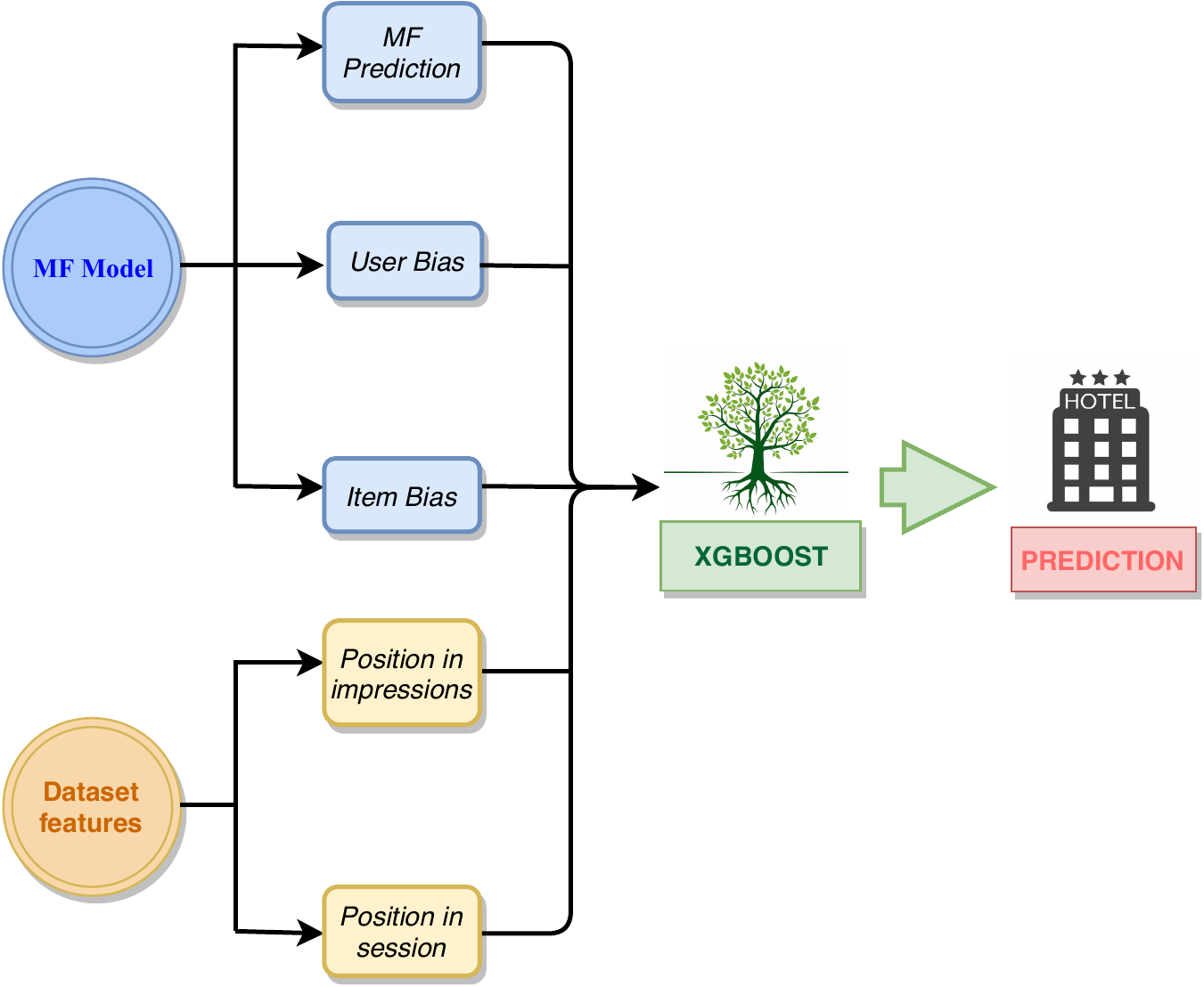}
    \caption{Matrix factorization with XGBoost approach.}
    \label{fig:mf_xgboost}
\end{figure}

\subsubsection{Gradient boosting}
\label{sec:mf_xgboost}

XGBoost is an optimized implementation of the gradient boosting algorithm~\cite{Chen2016}. Gradient boosting is a machine learning technique used to generate a better outcome, given more than one weak classifier. In our approach, it was useful for adding the contribution of two particular features of the dataset: the position of the hotel in the impression list, and the position in user's session (how recent is the interaction between user and hotel). This process is summarized in Figure~\ref{fig:mf_xgboost}. The importance of each feature, according to XGBoost, is outlined in Table~\ref{tab:features}.
The XGBoost training requires an additional split. To submit a solution we simply use our \textit{local training set} to train the MF model and generate predictions for the \textit{local validation}. Then, we use these predictions to train the XGBoost model and make the final ones on the \textit{challenge test set}.

\begin{table}
\begin{tabular}{@{}cl@{}}
\toprule
\textbf{Importance} & \textbf{Feature name} \\ \midrule
1 & MF Prediction \\
2 & Position in session \\
3 & User Bias \\
4 & Position in impressions \\
5 & Item Bias \\ \bottomrule
\end{tabular}
\caption{The importance of each feature in XGBoost for the MF architecture.}
\label{tab:features}
\end{table}

\subsubsection{Optimization}
\label{sec:mf_optimization}

In this section we report the result given by the MF without the gradient boosting.

We considered the following parameters:
\begin{itemize}
    \item Epochs: {150, 200, 300}
    \item Number of components: {200, 300}
    \item Learning rate: {0.01, 0.1, 0.2}
    \item Learning schedule: {adagrad, adadelta}
\end{itemize}

As loss function we choose \textit{warp-kos} because it significantly outperformed all the others. The results we obtained on the local validation set are reported in Table~\ref{tab:pure-mf}.
We can observe that the \textit{adagrad} learning schedule performs worse than the \textit{adadelta} one. The configuration that we selected is the best one reported.

\begin{table}
\begin{tabular}{@{}lllll@{}}
\toprule
\textbf{Epochs} & \textbf{Components} & \textbf{L. rate} & \textbf{L. schedule} & \textbf{MRR} \\ \midrule
200             & 300                 & 0.1                    & adadelta                   & 0.577164     \\
150             & 300                 & 0.1                    & adadelta                   & 0.577132     \\
300             & 300                 & 0.1                    & adadelta                   & 0.577080     \\
200             & 300                 & 0.2                    & adadelta                   & 0.577062     \\
200             & 200                 & 0.1                    & adadelta                   & 0.576162     \\
200             & 300                 & 0.01                   & adagrad                    & 0.431659    \\ \bottomrule
\end{tabular}
\caption{MF parameters tuning.}
\label{tab:pure-mf}
\end{table}

\subsection{Gated Recurrent Unit}
\label{sec:rnn}

\begin{figure}
    \includegraphics[width=\linewidth]{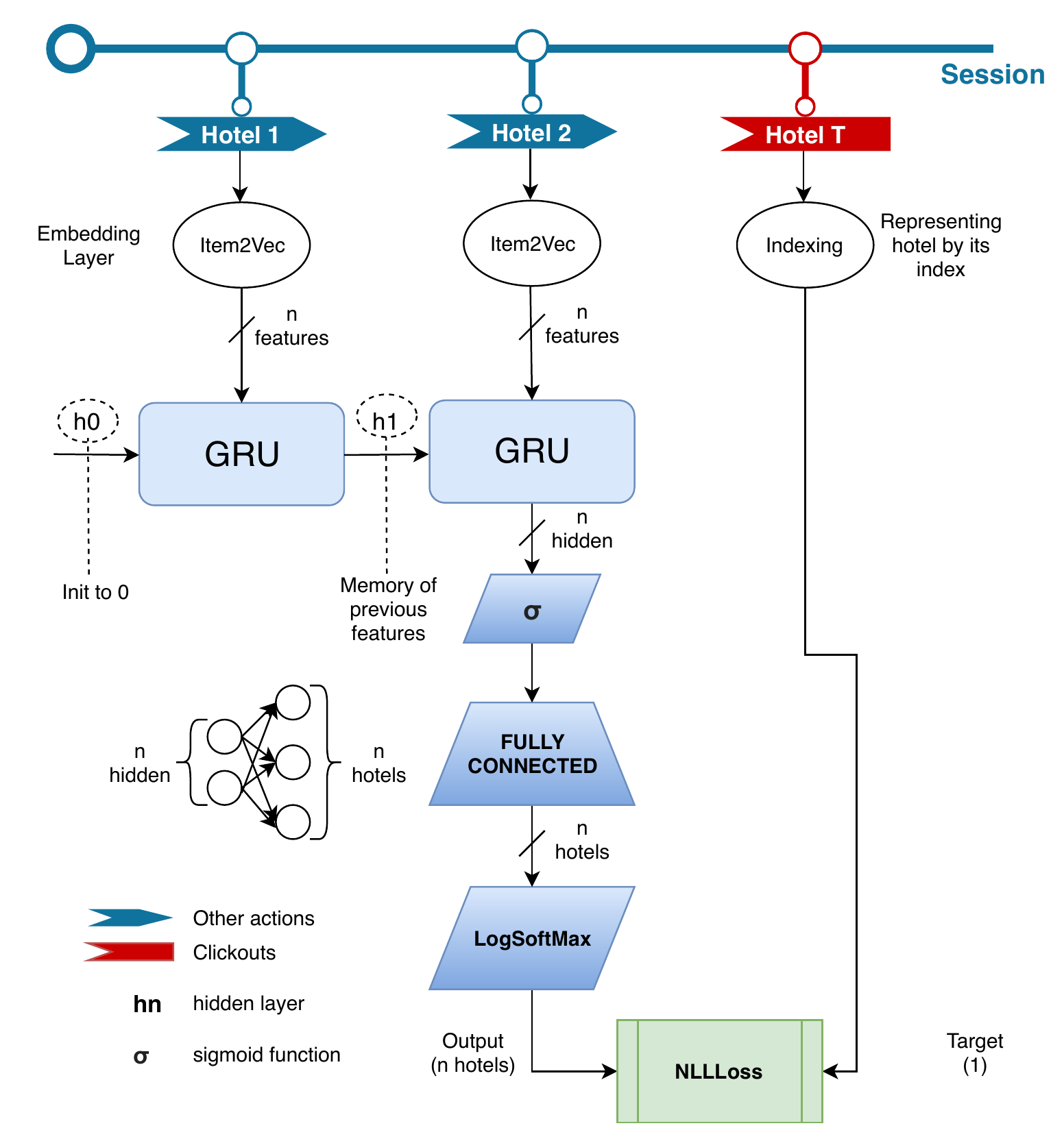}
    \caption{Recurrent Neural Network training architecture.}
    \label{fig:rnn}
\end{figure}

Recurrent Neural Network (RNN) is one of the most popular deep architectures~\cite{lecun2015deep}. It has been largely used in many fields, especially in Natural Language Processing for predicting the next word in a sentence~\cite{sutskever2011generating}. Its acknowledged ability in recognizing sequence patterns and in predicting the successive steps matches perfectly with the session-based problem we face, the only difference being in predicting just the final step of the sequence (\textit{clickout action}).

Specifically, we use gated Recurrent Unit (GRU) cells, discarding basic RNN for its gradient descent problem and LSTM for computation optimization~\cite{chung2014empirical}. The purpose is to represent the hotels (items) in a multi-dimensional space extracting their latent features, mapping similar hotels close to each other. In the following sections, we describe the input preparation, the training phase, and the prediction generation details.

\subsubsection{Item2Vec and Input Preparation}
\label{sec:rnn_input}

The dataset preparation phase consists of 3 steps: (1) remove every action unrelated to an item, as mentioned in Section~\ref{sec:approach}; (2) only keep the \textit{reference} field (hotel\_id) of the actions (our item); (3) keep the \textit{impression\_list} of null reference clickout actions for the prediction.

In order to extract correlation between different items, we use an approach based on word2vec embeddings~\cite{mikolov2013distributed}. We train the model by feeding it with a corpus composed of sentences (single sessions), each one composed by its items (hotel ids), collected from the whole dataset occurrences. More precisely, we used a model based on the Gensim implementation,\footnote{\url{https://radimrehurek.com/gensim/}} using the Skip-gram one with its default parameters: embedding vector dimension = 60, window size = 5, and min\_count = 1 for capturing every single item occurrence.

The difference in sequence length among sessions represents an issue.
The memory problem is solved by cutting the sessions to the last 200 actions, empirically observing that previous ones are not relevant for sequence purposes. The batch dimension problem is solved by padding every tensor in a batch to the length of the longest sessions in that batch.

The input $X$ is the collection of sessions, where each one is composed of a variable list of encoded (w2vec) items (hotels).

\begin{equation*}
X = {x^{S}_{0}, x^{S}_{1}, \dotsc, x^{S}_{n}}
\end{equation*}

\begin{equation*}
x_{i} = {h^{w}_{0}, h^{w}_{1}, \dotsc, h^{w}_{m}}
\end{equation*}

The target $Y$ is the list containing a single item target associated to each session, encoded by its index in the hotel list.

\begin{equation*}
Y = {y_{0}, y_{1}, \dotsc, y_{n}}
\end{equation*}

\subsubsection{Training and Network Architecture}
\label{sec:rnn_training}

As shown in Figure~\ref{fig:rnn}, the GRU cells carry previous item features by passing their hidden layer to the next step. Before the last step (clickout action), the last hidden layer is sent to a fully connected layer. This is required for expanding the latent features to get an output of size equal to the number of hotels, assigning a confidence score to every single existing item. We compare our network output with the hotel target (clickout reference) using a NLLLoss, optimal for classification purposes. We use a sigmoid function after the GRU for activation and we use a LogSoftMax before the loss function to get the scores in range (-inf, 0). We opt for NLLLoss instead of CrossEntropy because the former does not include the LogSoftMax layer, thus allowing to better analyze the scores for the prediction part.

\subsubsection{Prediction Generation}
\label{sec:rnn_prediction}

The strategy for obtaining the recommendation list begins by extracting and encoding all items in the session, feeding the neural network and obtaining the confidence score for every item, promoting those sharing similar latent features with the ones present in the session.

The second step is designed to extract the \textit{impression\_list} field from the clickout we want to predict, because it contains the list of items the user can choose from, which we want to order. Using a map, we associate every hotel in this list with its confidence score, thus allowing us to rank it by using a simple sort to obtain the final \textit{item\_recommendation} list.

\section{Experimental Results}
\label{sec:results}

The scores obtained on our local test set, as defined in Section~\ref{sec:approach}, are summarized in Table~\ref{tab:valscore}. 

\begin{table}
\begin{tabular}{lll}
\toprule
\textbf{Algorithm} & \textbf{MRR} \\ \midrule
MF + RNN           & 0.60277     \\
MF XGBoost         & 0.59804     \\
Pure RNN           & 0.28277     \\  \bottomrule
\end{tabular}
\caption{Local validation set results.}
\label{tab:valscore}
\end{table}

\begin{figure}
    \includegraphics[width=\linewidth]{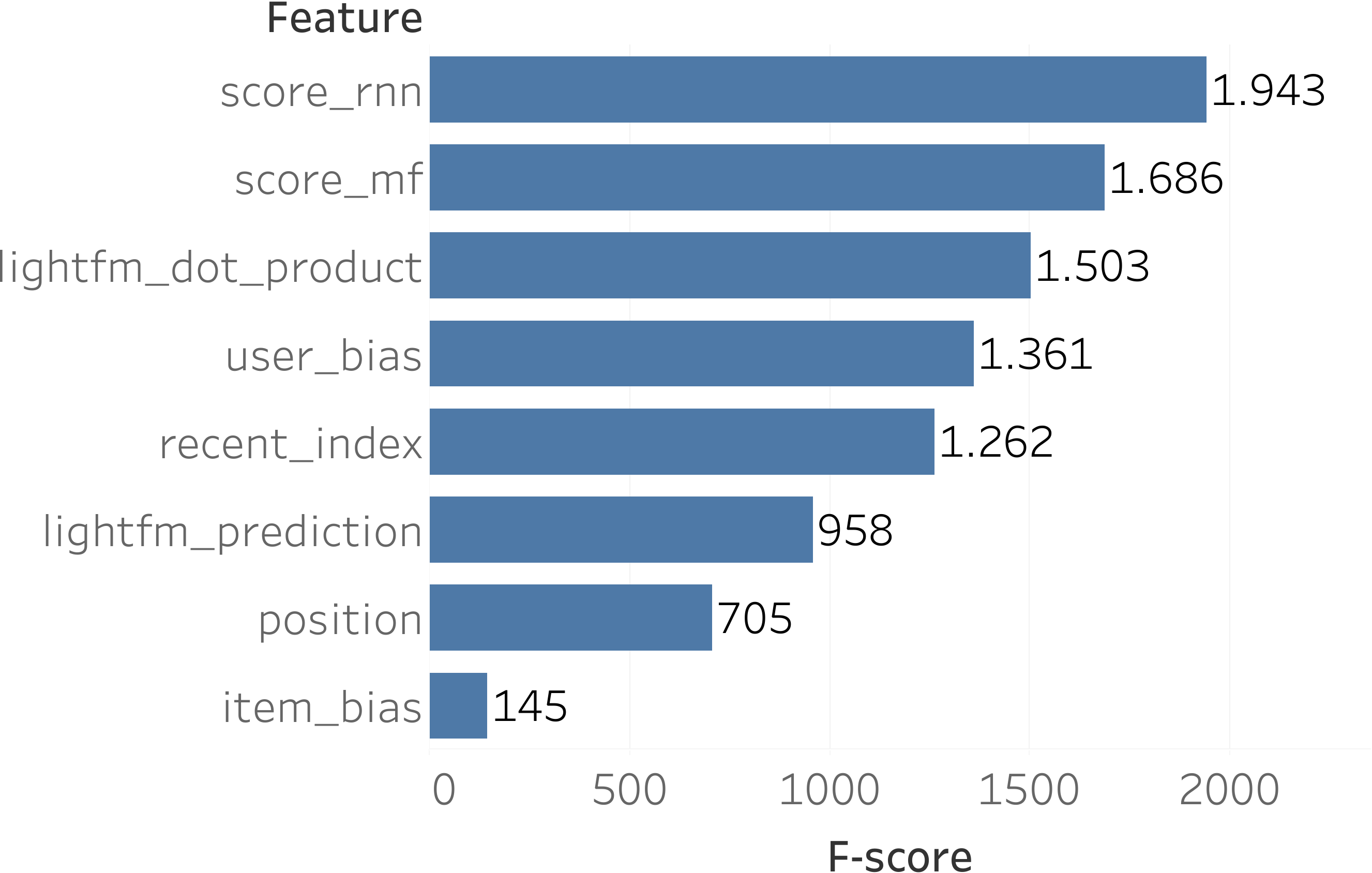}
    \caption{Importance of each feature in the XGBoost algorithm. The score represents the number of times a feature is used in the decision tree.}
    \label{fig:imp_xgb}
\end{figure}

The matrix factorization model obtains a score of \textbf{$0.5980$} on the validation set by the use of XGBoost on the features described in Section~\ref{sec:mf_xgboost}.

The score obtained by the RNN itself is 0.28277. It is not very high due to the huge data sparsity of the dataset, which penalizes the deep model. In the dataset the same hotel is present in very few sessions, making it difficult for the Item2Vec to find correlations between different hotels and leaving almost only the sequence of the different encoded items for the analysis.

The results obtained with the use of XGBoost for the ensemble are obtained by training it using the results of MF and RNN on the inner test set and then running it on their results over the local test set. The score obtained by the ensemble of the solutions is 0.602774 and it shows a slight improvement of the score.

In Figure~\ref{fig:imp_xgb}, the importance of each score shows that the most useful contributes to the decision of the solution are given by the main approaches, along with a smaller contribute of other features we also considered. The high contribution of the RNN compared to its low single-taken score suggests the successful extraction of latent features, important for the final score.

\section{Discussion and Ongoing Work}
\label{sec:conclusion}

In this paper, we presented the POLINKS solution to the RecSys Challenge 2019 based on an ensemble learning approach that uses the predictions generated by two different methods. 
The task of search ranking over a list of possible accommodations based on the session context and previous actions is not easy, especially when the number of features are limited.
Some important contextual features could not be used in the challenge, such as the number of people in the reservation, the stay duration as well as the travel date that captures the seasonality effect. These contextual features not only reflect the real nature of the traffic but can also affect a lot the ranking produced by a recommender system.

Matrix factorization is confirmed to be a solid tool for recommendations, even in a session-based problem lacking user history. The impact of the XGBoost optimization is huge: its amplification of multiple features contribute is a path to explore, even including other solution coefficients like the RNN ones. 

For what concerns the RNN, it does not get good results by itself for this particular dataset due to its sparsity, which negatively influences the Item2Vec and the deep architecture efficiency. The score may be improved with the development of a more suited encoding solution and stacking multiple deep architecture for different feature analysis.

The XGBoost ensemble of the two solutions shows a slight score improvement due to the complementarity of sequence analysis (RNN) and user-item interactions (MF), which is worth to be exploited in a competitive leaderboard such as the RecSys Challenge. Furthermore, a slight increase may become more important considering the margin of improvement of the RNN.

\bibliographystyle{ACM-Reference-Format}
\bibliography{references} 

\end{document}